# Revealing the charge density wave caused by Peierls instability in two-dimensional NbSe$_2$

Yung-Ting Lee[1,2], Po-Tuan Chen[3], Zheng-Hong Li[1], Jyun-Yu Wu[1], Chia-Nung Kuo[4,5], Chin-Shan Lue[4,5], Chien-Te Wu[1,6], Chien-Cheng Kuo[7], Cheng-Tien Chiang[8,9], Chun-Liang Lin[1]*, Chi-Cheng Lee[9,10]*, Hung-Chung Hsueh[9,10]*, Ming-Chiang Chung[6,11,12]*

[1]Department of Electrophysics, National Yang Ming Chiao Tung University; Hsinchu, 300093, Taiwan.
[2]Department of Chemistry, Tamkang University; Tamsui, New Taipei City, 251301, Taiwan.
[3]Department of Vehicle Engineering, National Taipei University of Technology; Taipei, 10608, Taiwan.
[4]Department of Physics, National Cheng Kung University; Tainan City, 70101, Taiwan.
[5]Taiwan Consortium of Emergent Crystalline Materials, National Science and Technology Council; Taipei, 10601, Taiwan.
[6]Physics Division, National Center for Theoretical Sciences; Taipei, 10617, Taiwan.
[7]Department of Physics, National Sun Yat-sen University; Kaohsiung, 80424, Taiwan.
[8]Institute of Atomic and Molecular Sciences, Academia Sinica; Taipei, 10617, Taiwan.
[9]Department of Physics, Tamkang University; Tamsui, New Taipei City, 251301, Taiwan.
[10]Research Center of X-ray Science, College of Science, Tamkang University; Tamsui, New Taipei City, 251301, Taiwan.
[11]Department of Physics, National Chung Hsing University; Taichung, 40227, Taiwan.
[12]Max Planck Institute for the Physics of Complex Systems; Nöthnitzer Straße 38, Dresden, 01187, Germany.

*Corresponding authors. Email: clin@nycu.edu.tw, cclee@mail.tku.edu.tw, hchsueh@gms.tku.edu.tw, mingchiangha@phys.nchu.edu.tw

**Abstract:** The formation of a charge density wave (CDW) in two-dimensional (2D) materials caused by Peierls instability is a controversial topic. This study investigates the extensively debated role of Fermi surface nesting in causing the CDW state in 2H-NbSe$_2$ materials. Four NbSe$_2$ structures (i.e., normal, stripe, filled, and hollow structures) are identified on the basis of the characteristics in scanning tunneling microscopy images and first-principles simulations. The calculations reveal that the filled phase corresponds to Peierls' description; that is, it exhibits fully opened gaps at the CDW Brillouin zone boundary, resulting in a drop at the Fermi level in the density of states and the scanning tunneling spectroscopy spectra. The electronic susceptibility and phonon instability in the normal phase indicate that the Fermi surface nesting is triggered by two nesting vectors, whereas the involvement of only one nesting vector leads to the stripe phase. This comprehensive study demonstrates that the filled phase of NbSe$_2$ can be categorized as a Peierls-instability-induced CDW in 2D systems.

**One-Sentence Summary:**

A charge density wave caused by Fermi-surface instability is demonstrated in NbSe$_2$ through a two-nesting-vector mechanism.

**Main Text:** The quantum phase transitions of a charge density wave (CDW) in two-dimensional (2D) materials have been extensively studied. CDWs are originally thought to be caused by the instability of the one-dimensional (1D) system described by Peierls (*1*); that is, Fermi surface nesting dictates the wave vector of the CDW ($Q_{CDW}$) and the corresponding lattice distortion. Fig. 1A shows the free-electron-like band of a 1D chain that has one electron per atomic site and an interatomic distance of *a*. The Fermi points are located at $k_F = \pm\pi/2a$ and connected by the nesting vector $Q_{CDW} = 2k_F$. Peierls asserted that this 1D system is unstable because of its divergent behavior in the response function at $Q_{CDW}$; this divergence indicates a possible electronic disturbance with the wave vector $2k_F$, which changes the periodicity of the chain and opens up a gap for gaining electronic energy near the Fermi level at the Brillouin zone (BZ) boundary of the new unit cell containing two atoms ($k = \pi/2a$; Fig. 1, B and C). In this 1D Peierls model, only a single $Q_{CDW}$ is required to couple Fermi points.

In a 2D system, Fermi surface nesting may also occur when a single $Q_{CDW}$ connects a large parallel portion of the Fermi surface. This phenomenon also indicates instability characterized by considerable electronic susceptibility, and a small perturbation can cause a charge density redistribution and/or Fermi surface reconstruction. When the electronic energy gain prevails over the Coulomb repulsion caused by a distorted structure, the CDW state can stabilize. In contrast to the 1D model, in the aforementioned 2D model, the Fermi surface need not be deployed at the new BZ boundary. On the other hand, in 2D free electron gas where no unique single $Q_{CDW}$ vector can be defined (see the band in the 2D hexagonal lattice in Fig. 1D), the Fermi surface can still be nested at a new BZ boundary to gain energy through gap openings (Fig. 1, E and F). Because none of the large portions of the Fermi surface is connected by a single $Q_{CDW}$, no divergent behavior is expected in the electronic susceptibility of the system. This subtle point is addressed in this study's discussion of the Fermi surfaces of $NbSe_2$ (Fig. 1, G to J). Originally, the CDW was assumed to exist only in 1D systems and to occur solely because of Peierls instability. Although the CDW state in 2D materials may still be related to Peierls' description, a growing body of theoretical and experimental evidence is indicating that this explanation does not apply to various real systems (*2*). For example, no relevant gap is observed at the Fermi level.

$NbSe_2$ is a typical 2D CDW material that has been extensively studied. The origin of the CDW state in $NbSe_2$ has been continuously debated (*3*). A first-principles density functional theory study on single-layer $NbSe_2$ with a commensurable 3 × 3 modulation revealed that multiple structures with nearly equal stability but different distortion patterns are compatible with a 3 × 3 modulation (*4*). This finding suggests the presence of a flat potential energy surface and the plausible coexistence of multiple modulations. Several structures were reported to coexist in a narrow energy range of 2−3 meV per atom (*5*). With respect to CDW phases, scanning tunneling microscopy (STM) was employed to monitor the transition of a $NbSe_2$ CDW between tridirectional (triangular) and unidirectional (stripe) structures (*6*,*7*). Subsequent theoretical and experimental studies have indicated that two structures can form triangular features in STM images (*8*–*10*); these two structures are referred to as filled and hollow structures in the present study (*8*). The relative stability of these two structures can be subtly altered by minuscule disturbances. Although these structures were regarded as CDW phases because of the newly developed periodicity, the data and modeling results pertaining to $NbSe_2$ CDWs indicate that Peierls instability is irrelevant in this context (*7*,*11*–*13*).

CDWs were assumed to be the result of electronic instability at the Fermi energy and the removal of this instability was through energy minimization involving periodic lattice modulation in a 1D Peierls model. A noteworthy topic is whether this description is valid and applicable in a real 2D system. First-principles calculation is a valuable tool for conducting theoretical studies of band structures and corresponding phase transitions. In the present study, we have investigated the CDW transition in NbSe$_2$ from theoretical and experimental perspectives and explored the mechanism of a CDW transition in 2D materials. An unanticipated discovery of the present study is that the formation of a filled structure exhibits the characteristics of a Peierls transition and that this CDW state is caused by two nesting vectors instead of a single $\mathbf{Q}_{CDW}$.

Four optimized structures are obtained through first-principles calculations, namely the normal, stripe, filled, and hollow phases (Fig. 2, A to D); the normal phase without a CDW is presented as a 3 × 3 supercell for consistency. In our theoretical tests, the monolayer result is consistent with that obtained through bulk calculations (see the supplemental material); thus, we focus on the monolayer case for our theoretical analysis. The four geometrical structures can be distinguished by patterns that comprised Se atoms and surrounding Nb–Nb bonds. These four structures correspond to the STM measurements (Fig. 2, E to H). Notably, in Fig. 2, E to H, the filled and hollow structures form triangular features. The various patterns displayed in the STM images can be explained through calculated charge density. The charge distribution in the filled CDW (Fig. 2C) indicates that one Nb atom has a charge that is different from those of other Nb atoms, resulting in an exceptionally bright spot at the corner of the unit cell in the STM image (Fig. 2G). In contrast to the filled structure, the hollow structure has not one but three Nb atoms with an equal charge (Fig. 2D); thus, three bright spots are observed in the corresponding STM image (Fig. 2H). Among the three nonnormal phases, the filled phase exhibits the lowest energy level (ΔE = −0.96 meV/atom) relative to that of the normal phase in the 3 × 3 × 1 supercell. Further, our calculations indicate that the relative energy levels of the filled phase and the hollow phase (ΔE = −0.81 meV/atom) differed by only 0.15 meV. This minuscule difference explains the coexistence of the filled and hollow structures, which was observed in previous STM experiments (*9*).

NbSe$_2$ CDW states were believed to be unrelated to Peierls instability, which requires a gap opening at the Fermi level. To clarify the related causal relationship, the electronic band structures and Fermi surfaces of the normal, stripe, filled, and hollow phases are presented in Fig. 3, A to D, respectively. The gaps are fully open at the supercell BZ boundary in the filled phase but not in the stripe and hollow phases. This has not been revealed in previous studies (*9, 14*); specifically, the CDW of NbSe$_2$ was not reported to exhibit an energy gap at the Fermi level. In the hollow phase, the gap opening forms below the Fermi level, which is consistent with the findings of another study (*14*). The Fermi surface of the stripe phase resembles that of the filled phase along one specific direction; this finding will be discussed in a subsequent paragraph. In the filled phase, the feature of the fully opened gap of the Fermi surface at the CDW BZ boundary manifests the effect of Fermi surface nesting.

Through first-principles calculations, we obtain a favorably converged electronic density of states (DOS), enabling us to determine the location of the DOS drop and its shape near the Fermi level in the freestanding NbSe$_2$ monolayer phases. Fig. 3E reveals that only the filled phase undergoes a prominent drop at the Fermi level and exhibits a slightly asymmetric feature because of the lack of particle–hole symmetry in its band structure; this finding is also reflected in its scanning tunneling spectroscopy spectra (Fig. 3F). The gap range in the measured spectra differs from that of the DOS because multiple coexisting structures are present in our samples. Although drops can also be observed in the hollow and stripe phases, the locations of these drops deviated from the Fermi level; that is, they do not correspond to Peierls' description. The total energy calculations also indicate that the largest energy gain occurs in the filled phase. On the basis of the calculated band structure, Fermi surface, and DOS, the formation of the filled phase is revealed as a Peierls transition.

The relevant nesting vectors are examined on the basis of electronic susceptibility, and the matrix elements are set to unity for first-principles calculations (*15*). The peaks in the imaginary part of the primitive-cell response function of the normal phase are shown in Fig. 3G, and they reveal the strong tendency of a CDW to form through the single nesting vector $\mathbf{Q}_K$: (1/3, 1/3, 0) in the units of the reciprocal lattice vectors, which were revealed by the high intensity in Fig. 1G. However, $\mathbf{Q}_K$ is inefficient in producing a stable CDW state because the hole pockets at Γ and K can not be matched perfectly, and a direct folding, which occurs through $\mathbf{Q}_K$ in a $\sqrt{3} \times \sqrt{3}$ supercell with a 30° rotation, can not deploy a Fermi surface at the supercell boundary (Fig. 1H). This inefficiency is consistent with the property of the real part of the response function; instead of a divergent peak at K, only a shallow peak is detected at approximately $\mathbf{Q}_{2M/3} \cong (2/3)$ [ΓM] along Γ to M: (0.5, 0, 0) (Fig. 3H). Therefore, the formation of the 3 × 3 filled phase, which is also commensurate with $\mathbf{Q}_K$, can not be directly triggered by $\mathbf{Q}_{CDW} = \mathbf{Q}_K$. In 2D, $\mathbf{Q}_K$ is equal to $\mathbf{Q}_{2M/3} + \mathbf{Q}_{2M'/3}$ with M´: (0, 0.5, 0), and all symmetry-related $\mathbf{Q}_{2M/3}$ and $\mathbf{Q}_K$ points can meet at the hexagonal BZ center of a 3 × 3 supercell, enabling all the pockets to surround the BZ boundary (Fig. 1I) to form fully opened gaps at the BZ boundary (Fig. 1J). Therefore, the fully opened gaps driven by Fermi surface nesting in the filled phase are inferred to originate from the combination of $\mathbf{Q}_{2M/3}$ and $\mathbf{Q}_{2M'/3}$.

$\mathbf{Q}_{2M/3}$ is closely related to the formation of the stripe phase as shown in Fig. 2, B and F, which is presented using a 3 × 3 supercell; however, the primitive cell is essentially 3 × 1 in terms of its consistency with the periodicity dictated by the $\mathbf{Q}_{2M/3}$ vector. This finding also explains why the Fermi surfaces of the filled and stripe phases are partially similar. When Fermi surface nesting dominates CDW formation, at least one vibrational mode must undergo softening at low temperature to drive lattice distortion. Accordingly, soft modes with imaginary frequencies are detected in the normal phase and appear in the primitive BZ (Fig. 4A). The most unstable phonon mode is also revealed to be located at $\mathbf{Q}_{2M/3}$, and it is referred to as the $\mathbf{Q}_{2M/3}$-mode. After the atoms are moved in accordance with the eigen-displacements of the $\mathbf{Q}_{2M/3}$-mode (Fig. 4B), the structural phase transition from the normal phase to the stripe phase can be reproduced. Our calculations verify that the stripe phase, which has the softest mode (Fig. 4C), is still dynamically unstable because of the remaining instability at $\mathbf{Q}_{2M'/3}$ (Fig. 4D). Guided by the eigen-displacements of the $\mathbf{Q}_{2M'/3}$-mode, the filled structure can be reached, and the phonon dispersion in the filled phase (Fig. 4E) indicates its dynamical stability. In the hollow phase, no gaps are opened at the Fermi level at the CDW boundary; however, the phase is also dynamically stable (Fig. 4F).

The aforementioned findings indicate that the stripe phase is an essential mediator of the normal-to-filled phase transition, and two nesting vectors, $\mathbf{Q}_{2M/3}$: (1/3, 0, 0) and $\mathbf{Q}_{2M'/3}$: (0, 1/3, 0), are required to cause Fermi surface nesting with fully opened gaps at the hexagonal BZ boundary. These two nesting vectors can be revealed through electronic susceptibility and phonon instability. The collected evidence indicates that the Peierls-instability-driven CDW state can occur in real 2D materials.

**References**

(1) R. E. Peierls, Quantum Theory of Solids (Oxford University Press, New York) (1955).
(2) M. D. Johannes, I. I. Mazin, Fermi surface nesting and the origin of charge density waves in metals. *Phys. Rev. B* **77** (16), 165135 (2008).
(3) X. Zhu, Y. Cao, J. Zhang, E. W. Plummer, J. Guo, Classification of charge density waves based on their nature. *Proc. Natl. Acad. Sci. USA* **112** (8), 2367 (2015).
(4) J. A. Silva-Guillén, P. Ordejón, F. Guinea, E. Canadell, Electronic structure of 2H-NbSe$_2$ single-layers in the CDW state. *2D Mater* **3**, 035028 (2016).

(5) B. Guster, C. Rubio-Verdú, R. Robles, J. Zaldívar, P. Dreher, M. Pruneda, J. Á. Silva-Guillén, D.-J. Choi, J. I. Pascual. M. M. Ugeda, P. Ordejón, E. Canadell, Coexistence of Elastic Modulations in the Charge Density Wave State of 2H-NbSe$_2$. *Nano Lett.* **19** (5), 3027−3032 (2019).
(6) F. Bischoff, W. Auwärter, J. V. Barth, A. Schiffrin, M. Fuhrer, B. Weber, Nanoscale Phase Engineering of Niobium Diselenide. *Chem. Mater.* **29** (23), 9907−9914 (2017).
(7) A. Soumyanarayanan, M. M. Yee, Y. He, J. van Wezel, D. J. Rahn, K. Rossnagel, E. W. Hudson, M. R. Norman, J. E. Hoffman. Quantum phase transition from triangular to stripe charge order in NbSe$_2$. *Proc. Natl. Acad Sci. USA* **110** (5), 1623–1627 (2013).
(8) C.-S. Lian, C. Si, W. Duan, Unveiling Charge-Density Wave, Superconductivity, and Their Competitive Nature in Two-Dimensional NbSe$_2$. *Nano Lett.* **18** (5), 2924 (2018).
(9) A. Sanna, C. Pellegrini, E. Liebhaber, K. Rossnagel, K. J. Franke, E. K. U. Gross, Real-space anisotropy of the superconducting gap in the charge-density wave material 2H-NbSe$_2$. *npj Quantum Materials* **7**, 6 (2022).
(10) S. Divilov, W. Wan, P. Dreher, E. Bölen, D. Sánchez-Portal, M. M. Ugeda, F. Ynduráin, Magnetic correlations in single-layer NbSe$_2$. *J. Phys.: Condens. Matter* **33** (29), 295804 (2021).
(11) X. Zhu, Y. Cao, J. Zhang, E. W. Plummer, J. Guo, Classification of charge density waves based on their nature. *Proc. Natl. Acad. Sci. USA* **112** (8), 2367–2371 (2015).
(12) M. Calandra, I. I. Mazin, F. Mauri, Effect of dimensionality on the charge-density wave in few-layer 2H-NbSe$_2$. *Phys. Rev. B* **80** (24), 241108 (2009).
(13) F. Weber, S. Rosenkranz, J.-P. Castellan, R. Osborn, R. Hott, R. Heid, K.-P. Bohnen, T. Egami, A. H. Said, D. Reznik, Extended phonon collapse and the origin of the charge-density wave in 2H-NbSe$_2$. *Phys. Rev. Lett.* **107** (10), 107403 (2011).
(14) C. J. Arguello, S. P. Chockalingam, E. P. Rosenthal, L. Zhao, C. Gutiérrez, J. H. Kang, W. C. Chung, R. M. Fernandes, S. Jia, A. J. Millis, R. J. Cava, A. N. Pasupathy, Visualizing the charge density wave transition in 2H-NbSe$_2$ in real space. *Phys. Rev. B* **89** (23), 235115 (2014).
(15) M. D. Johannes, I. I. Mazin, C. A. Howells, Fermi-surface nesting and the origin of the charge-density wave in NbSe$_2$. *Phys. Rev. B* **73** (20), 205102 (2006).

## Acknowledgments

The research study is supported financially by the Ministry of Science and Technology (MOST) of Taiwan (Grant Nos. 110-2112-M-A49-022-MY2, 111-2112-M-005-013, and 110-2112-M-032-014-MY3).

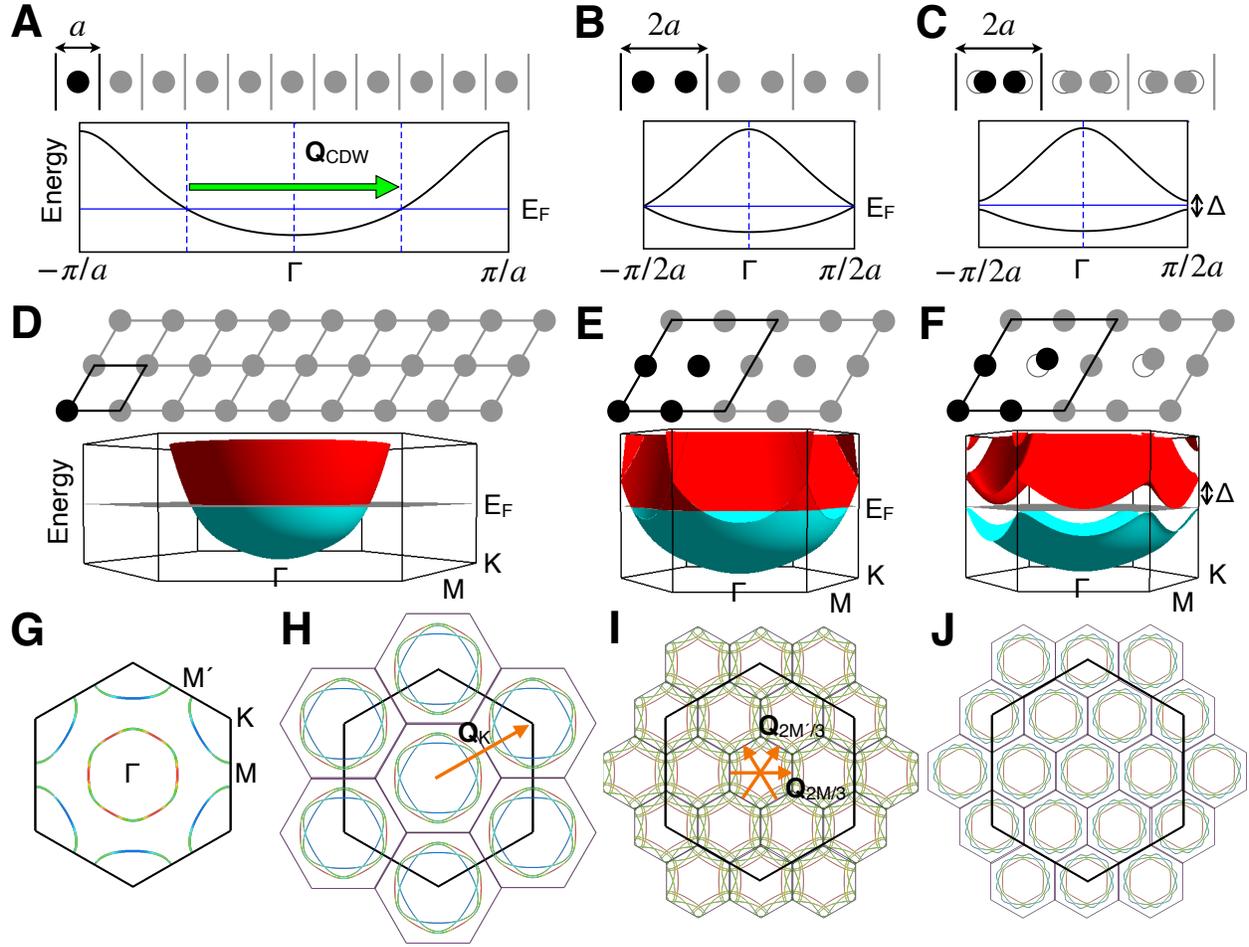

**Fig. 1. Illustration of Peierls' description of charge density wave (CDW).** Band structures of a 1D chain corresponds to (**A**) an atom per unit cell, (**B**) two atoms per perfect supercell, and (**C**) two atoms per supercell involving displaced atoms. The new periodicity guided by the Fermi point nesting vector ($Q_{CDW}$ in (A)) opens the gap ($\Delta$ in (C)) and helps gain the electronic energy. (**D** to **F**) Peierls CDW extended to a 2D hexagonal system. Fermi surfaces of normal-phase NbSe$_2$ monolayer presented in the Brillouin zone of (**G**) a primitive cell, (**H**) a $\sqrt{3} \times \sqrt{3}$ supercell, and (**I**) a $3 \times 3$ supercell. (**J**) The Fermi surface of the filled-phase NbSe$_2$ monolayer with the $3 \times 3$ supercell. Fermi velocities are colored blue (low) and red (high) from in (G to J).

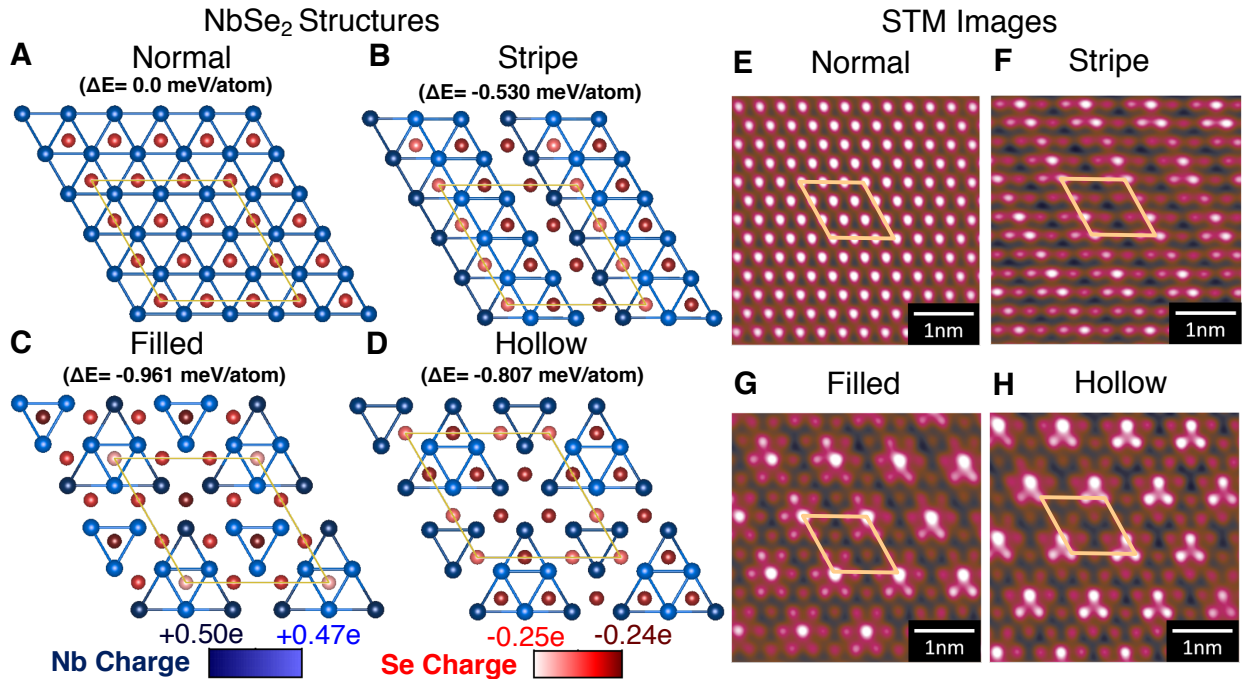

**Fig. 2. Structure, charge distribution, and STM image of NbSe$_2$ CDW phases.** Structure and charge distribution of (**A**) normal, (**B**) stripe, (**C**) filled, and (**D**) hollow phases from calculations. Total energy per atom in the normal phase with 3 × 3 × 1 supercell serves as a reference for evaluating relative energy difference. Different atomic charge states are indicated by different colors. Experimental scanning tunneling microscopy (STM) images of (**E**) normal, (**F**) filled, (**G**) hollow, and (**H**) stripe phases. The 3 × 3 × 1 supercells are shown as orange rhombuses.

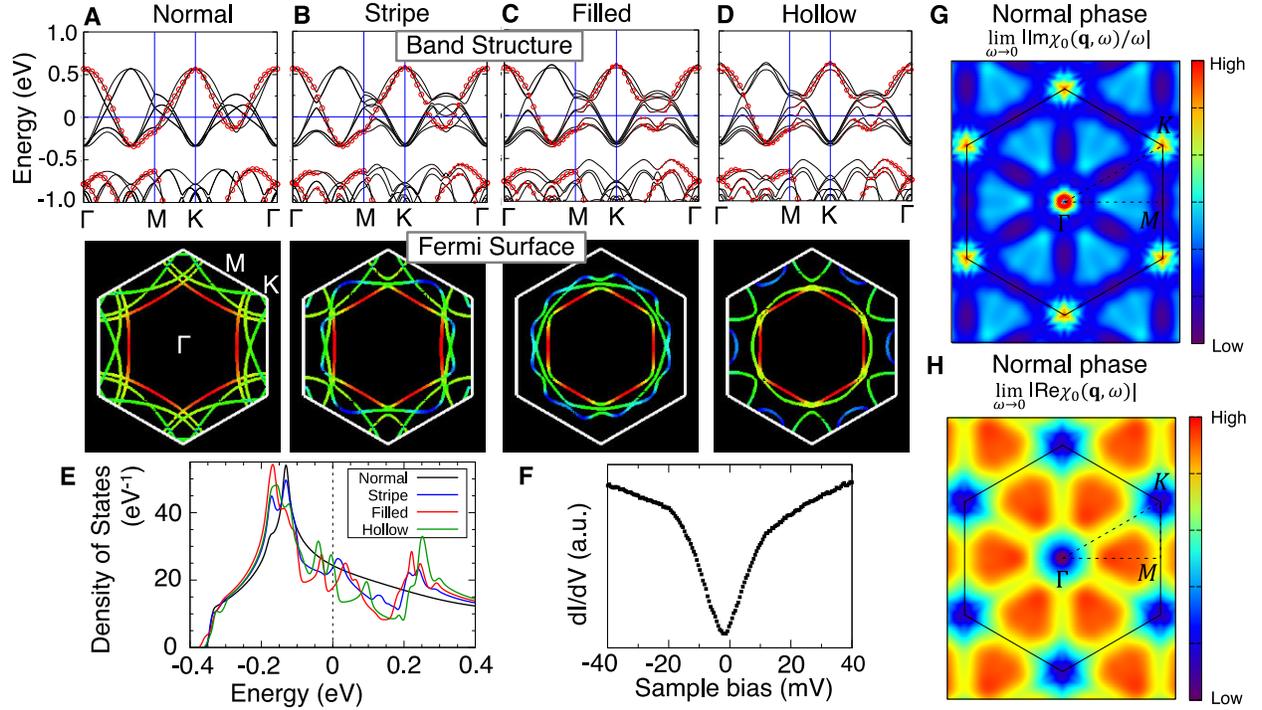

**Fig. 3. Electronic structure and susceptibility of NbSe$_2$.** Band structure (up) and Fermi surface (down) of (**A**) normal, (**B**) stripe, (**C**) filled, and (**D**) hollow phases, respectively. All band structure calculations pertaining to 3 × 3 supercell (black curves) are unfolded to Brillouin zone of (1 × 1) primitive cell (red circles). Fermi velocity is indicated by color (A to D) in lower panels. (**E**) Density of states per supercell. For consistency, density of states of normal phase is also calculated using 3 × 3 supercell. Fermi level is set at 0 eV. (**F**) Experimental scanning tunneling spectroscopy spectrum of NbSe$_2$. Imaginary and real parts of electronic susceptibility of normal phase are shown in (**G**) and (**H**).

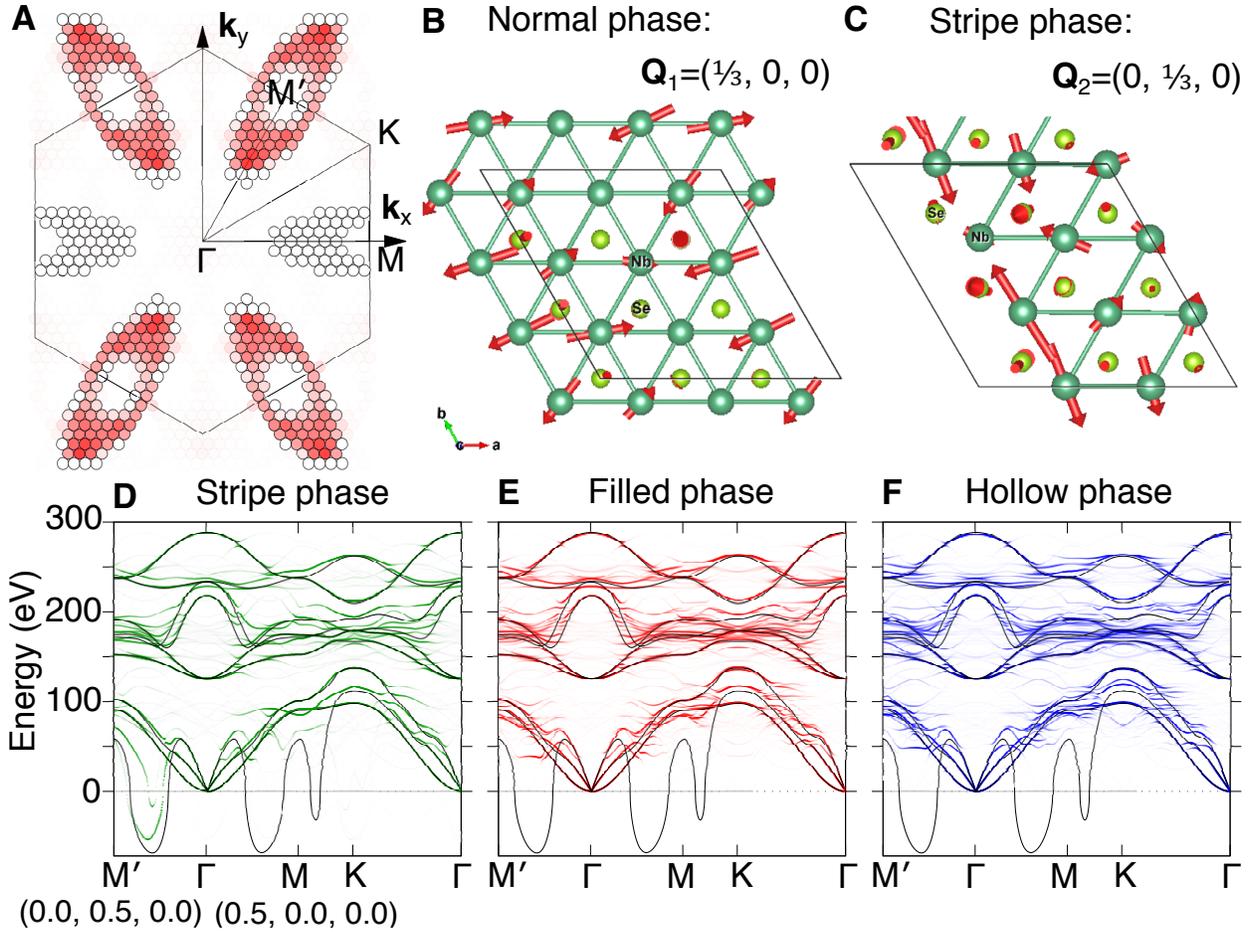

**Fig. 4. Phonon dispersion in NbSe$_2$ CDW phases.** (**A**) Locations of the imaginary frequencies of phonon dispersion in normal and stripe phases at ($k_x$, $k_y$) plane are indicated by black and red circles in the Brillouin zone, respectively. Eigen displacement of imaginary frequency of (**B**) normal phase at $Q_1 = (1/3,0,0)$ and (**C**) stripe phase at $Q_2 = (0,1/3,0)$. Phonon dispersion of (**D**) stripe phase, (**E**) filled phase, and (**F**) hollow phase. To ease the comparison of the phonon dispersion in (D) stripe, (E) filled, and (F) hollow phases with the normal-phase dispersion (black curves), the dispersions with the unfolded weights to the primitive-cell Brillouin zone of the normal phase are presented by green, red, and blue circles, respectively.